\documentstyle[twoside,epsfig,fleqn,espcrc2]{article}

\newcommand{\AmS}{{\protect\the\textfont2
  A\kern-.1667em\lower.5ex\hbox{M}\kern-.125emS}}

\title{Measurement of the photon structure function at ALEPH}

\author{A.~B\"ohrer\address{Fachbereich Physik, University of Siegen,\\
        57068 Siegen, Germany} \hfill {Si-99-03}}
       
\begin{document}

\begin{abstract}
The photon structure function $F_2^{\gamma}$ has been measured with data 
taken by the ALEPH collaboration at LEP centre-of-mass energies 
$\sqrt{s} = 91\,{\mathrm{GeV}}$ with $\langle Q^2 \rangle$ of 
9.9, 20.7 and $284\,{\mathrm{GeV^2}}$ and 
$\sqrt{s} = 183\,{\mathrm{GeV}}$ with $\langle Q^2 \rangle$ of 
13.7 and $56.5\,{\mathrm{GeV^2}}$. For the 
data at $\sqrt{s} = 183\,{\mathrm{GeV}}$ a two-dimensional unfolding method 
employing the principle of 
maximum entropy is used, which reduces the errors compared to 
one-dimensional methods.
\end{abstract}

\maketitle

\section{INTRODUCTION}

The photon structure function $F_2^{\gamma}$ in 
${{\mathrm e}^+{\mathrm e}^-}$ collisions is measured 
\cite{f2gamma91,f2gamma183} where one of the 
incident beam leptons is scattered at sufficiently large angle to be 
detected. In these singly-tagged events the negative momentum transfer 
$Q^2 = - q^2$ of the photon emitted by the tagged lepton is 
$Q^2 = 2 E_{\mathrm{tag}} E_{\mathrm{beam}} (1-\cos \theta)$. The second 
lepton stays undetected so the momentum transfer is small. 

The process then can be viewed as inelastic electron-photon scattering, 
where a quasi-real photon is probed by a virtual photon \cite{berger}:

\begin{eqnarray}
\frac{ {\mathrm d^2}\sigma_{{\mathrm e}\gamma \rightarrow {\mathrm {eX}}} }
{ {\mathrm d}x{\mathrm d}Q^2 } &=&
\frac{2\pi \alpha^2}{xQ^2} \left[ \left( 1+ (1-y)^2 \right)
F_2^{\gamma}(x,Q^2) \right. \nonumber \\
~&~& \left. {} - y^2 F_{L}^{\gamma}(x,Q^2) \right]
{\label{equ-f2}}
\end{eqnarray}
with
\begin{eqnarray}
x &\approx& \frac{Q^2}{Q^2+W^2} \\
y &\approx& 1 - \frac{E_{\mathrm {tag}}}{E}
         \cos ^2 \left( \theta_{\mathrm {tag}} \right)
\end{eqnarray}

The inelasticity, measured with $y$, is small and the term with $F_L^{\gamma}$ 
may be neglected. Equation \ref{equ-f2} can be used to relate the distribution 
in $x$ and $Q^2$ to the structure function $F_2^{\gamma}$. A more elaborate 
introduction and discussion of $F_2^{\gamma}$ you may find in 
Ref.~\cite{nisius}.

In this article the recent measurements of $F_2^{\gamma}$ with the 
ALEPH experiment are presented \cite{f2gamma91,f2gamma183}. Measurements 
at two different centre-of-mass energies and various $Q^2$ have been 
performed. The data are used to test different model parameterizations 
of $F_2^{\gamma}$. Special 
emphasis has been put on a two-dimensional unfolding method to extract
the true $x$ distribution ${\mathrm d}N/{\mathrm d}x$ from the measured data.

\section{DATA}

The ALEPH detector and its performance have been described in detail 
elsewhere \cite{aleph-det}. Charged tracks and neutral calorimeter 
energy as defined by the ALEPH energy flow package \cite{eflow} are used 
in these analyses.

Single tag events are selected by the lepton detected in the electromagnetic 
calorimeters of ALEPH: ECAL and LCAL for $\sqrt{s} = 91\,{\mathrm{GeV}}$ 
(LCAL and SiCAL for $\sqrt{s} = 183\,{\mathrm{GeV}}$). The scattering 
angle is measured in the range of 
$-0.6 < \cos \theta_{\mathrm {tag}} < 0.95$ in ECAL and 
$65\,{\mathrm{mrad}} < \theta_{\mathrm {tag}} < 150\,{\mathrm{mrad}}$ 
in LCAL ($60\,{\mathrm{mrad}} < \theta_{\mathrm {tag}} < 155 
\,{\mathrm{mrad}}$ at $183\,{\mathrm{GeV}}$). Further cuts on 
$E_{\mathrm {tag}}$ and a veto on a second tag are applied.

The visible hadronic final state has to consist of at least three 
charged tracks and has to have an invariant mass $W_{\mathrm {vis}} > 
2\,{\mathrm{GeV}}$. Further cuts for rejection of beam-gas events are applied. 
Additional cuts are required for the data at 
$\sqrt{s} = 91\,{\mathrm{GeV}}$ rejecting background from Z decays.

We are left with clean data samples with a remaining background of 
a few percent. The samples are divided into 
three (two) subsamples; see Table \ref{tab-samples}.

\begin{table}[hbt]
\caption{Data samples with their statistics and $Q^2$}
\label{tab-samples}
\begin{tabular}{|l|r|r|r|r|}
\hline
     & \# of    & $Q^2$ range           & $\langle Q^2 \rangle$ 
                                                 & $E_{\mathrm {cms}}$\\
     & events   & GeV$^2$               & GeV$^2$   & GeV \\
\hline
ECAL &    163   & 35 - 3000             & 284        &  91 \\
LCAL &   1647   & 13 - \phantom{00}44   & 20.67      &  91 \\
LCAL &   1543   &  6 - \phantom{00}13   & 9.93       &  91 \\
\hline
SiCAL &   1208   &  7 - \phantom{0}24   & 13.7       &  183\\
LCAL  &   861   & 17 - 200             & 56.5        &  183 \\
\hline
\end{tabular}
\end{table}

\begin{figure}[htb]
\vspace*{-0.6cm}
\epsfig{file=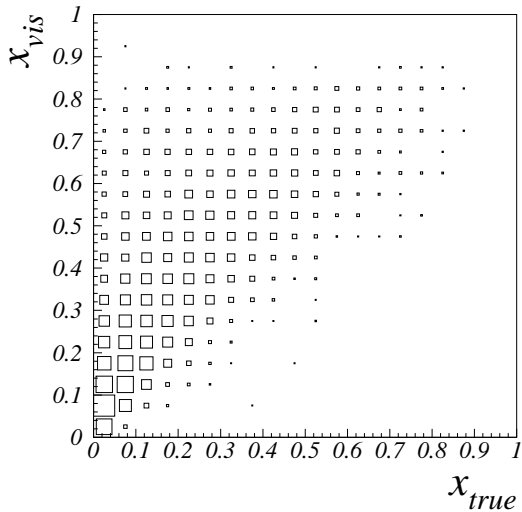,width=7.5cm}
\vspace*{-1.0cm}
\caption{$x_{\mathrm{vis}}$ versus $x_{\mathrm{true}}$ for LCAL 
tagged events at $\sqrt{s} = 91\,{\mathrm{GeV}}$}
\label{fig-lcal_xvis_xtrue}
\end{figure}

The data are corrected for trigger efficiency (which is close to 100\%), 
downscaling, and acceptance. For the correction it is important to note 
that the invariant mass $W$ and therefore $x$ is poorly measured only, 
see Fig.~\ref{fig-lcal_xvis_xtrue}. This 
is especially due to the significant portion of the energy of the event 
which goes in the forward region where there is little or no tracking 
information. Therefore extraction of a measurement of the photon 
structure function from the data depends on the use of a model of the 
production of hadronic final states from $\gamma^* \gamma$ collisions. There 
is at present no complete theoretical description of this process, so a 
number of phenomenological models are used.

Models giving reasonable descriptions of global event variables are then used 
to correct the data using an unfolding method. 

The models chosen at 
$\sqrt{s} = 91\,{\mathrm{GeV}}$ are firstly a combination of QPM and 
VDM \cite{berger,qpmvdm}, both with JETSET fragmentation \cite{jetset},  
and secondly HERWIG \cite{herwig} with GRV LO \cite{grv}. QPM and 
VDM are combined 
to form a single set, weighting each sample so as to give the best 
$\chi^2$ between distributions (the number of energy flow objects, the 
transverse momentum of all energy flow objects with respect to the 
beam direction, and the thrust of the event) 
predicted by the combined simulation and the data. 
In the HERWIG program the defaults for version 5.9 are used, apart from 
the intrinsic transverse momentum $k_{\mathrm t}$ of the partons in 
the target photon. This was modified according to the scheme proposed by ZEUS 
and described in Ref.~\cite{cartwright,zeus}.

The models chosen at
$\sqrt{s} = 183\,{\mathrm{GeV}}$ are firstly HERWIG \cite{herwig} with GRV 
LO \cite{grv}, secondly HERWIG \cite{herwig} with SaS \cite{sas} and thirdly 
PHOJET \cite{engel} (only used for SiCAL data, because the description of the 
LCAL data is only moderate).

\section{UNFOLDING AND \\ EXTRACTION OF $F_2^{\gamma}$ }

As pointed out in the previous section, the invariant mass $W$ is poorly 
measured by current detectors and an unfolding method has to be employed. 
Both at $\sqrt{s} = 91\,{\mathrm{GeV}}$ and at $183\,{\mathrm{GeV}}$ a 
regularization procedure is used to suppress oscillations in the 
result. 

At $\sqrt{s} = 91\,{\mathrm{GeV}}$ the unfolding was performed using the 
procedure proposed by Blobel \cite{blobel}. This procedure fits a sum of 
spline curves to the data after passing them through the $x_{\mathrm {vis}}$ 
versus $x_{\mathrm {true}}$ response matrix obtained from the simulated events 
(HERWIG and QPM+VDM), suppressing oscillations which have higher frequency. 
The structure function $F_2^{\gamma}$ is then obtained using the 
GALUGA program \cite{galuga}, where $F_L^{\gamma}$ is set to its 
theoretical value. With $F_2^{\gamma}$ set to 1 the output is used as 
reference for the extraction of $F_2^{\gamma}$ of the data.

\begin{figure}[htb]
\hspace*{-1.0cm}
\vspace*{-1.0cm}
\epsfig{file=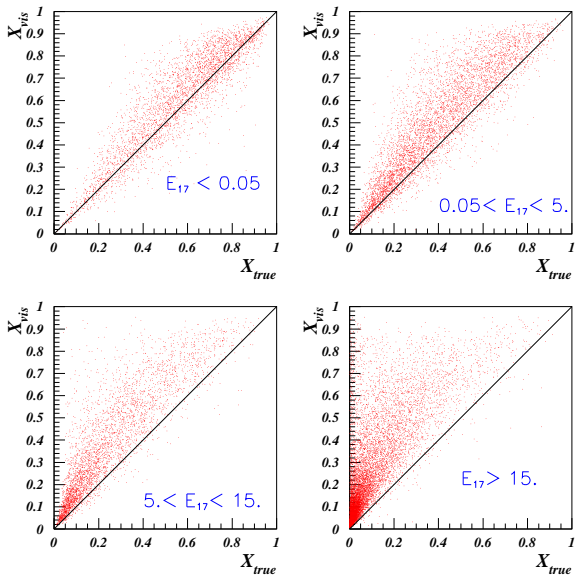,width=9.7cm}
\vspace*{-1.0cm}
\caption{$x_{\mathrm{vis}}$ versus $x_{\mathrm{true}}$ for subsequent bins
of the variable $E_{17}$ using HERWIG simulation.}
\label{fig-e17_res}
\end{figure}

At $\sqrt{s} = 183\,{\mathrm{GeV}}$ a new, recently proposed method 
\cite{cartwright} has been used for the unfolding (Ref.~\cite{f2gamma183} 
and references therein). It uses the principle 
of maximum entropy. In addition, this method allows a two-dimensional 
unfolding to be applied. The regularization function $S(\vec{\mu})$ used 
is the Shannon entropy \cite{shannon}:

\begin{equation}
S(\vec{\mu}) = - \sum_{j=1}^{M} \frac{\mu_j}{\mu_{\mathrm{tot}}} \log 
\left( \frac{\mu_j}{\mu_{\mathrm{tot}}} \right) \,
\end{equation}

when $\mu_j$ are the parameters to be estimated in bin $j$ and 
$\mu_{\mathrm{tot}}$ is their sum. This entropy-based regularization 
makes no reference to the relative locations of any of the bins. 
Therefore the principle of maximum entropy can be directly applied to 
multidimensional distributions. This method reduces the model dependence 
by including for each event not only $x$, but in addition some 
other variable characterizing the final state. For this, the variable $E_{17}$ 
has been introduced, defined as the total energy of the particles having 
angles with respect to the beam line less than $17^{\circ}$. Here 
the $x$ resolution degrades considerably for increasing $E_{17}$, see 
Fig.~\ref{fig-e17_res}.

\vspace*{-0.6cm}
\vspace*{-1.2cm}

\begin{figure}[htb]
\hspace*{-0.55cm}
\epsfig{file=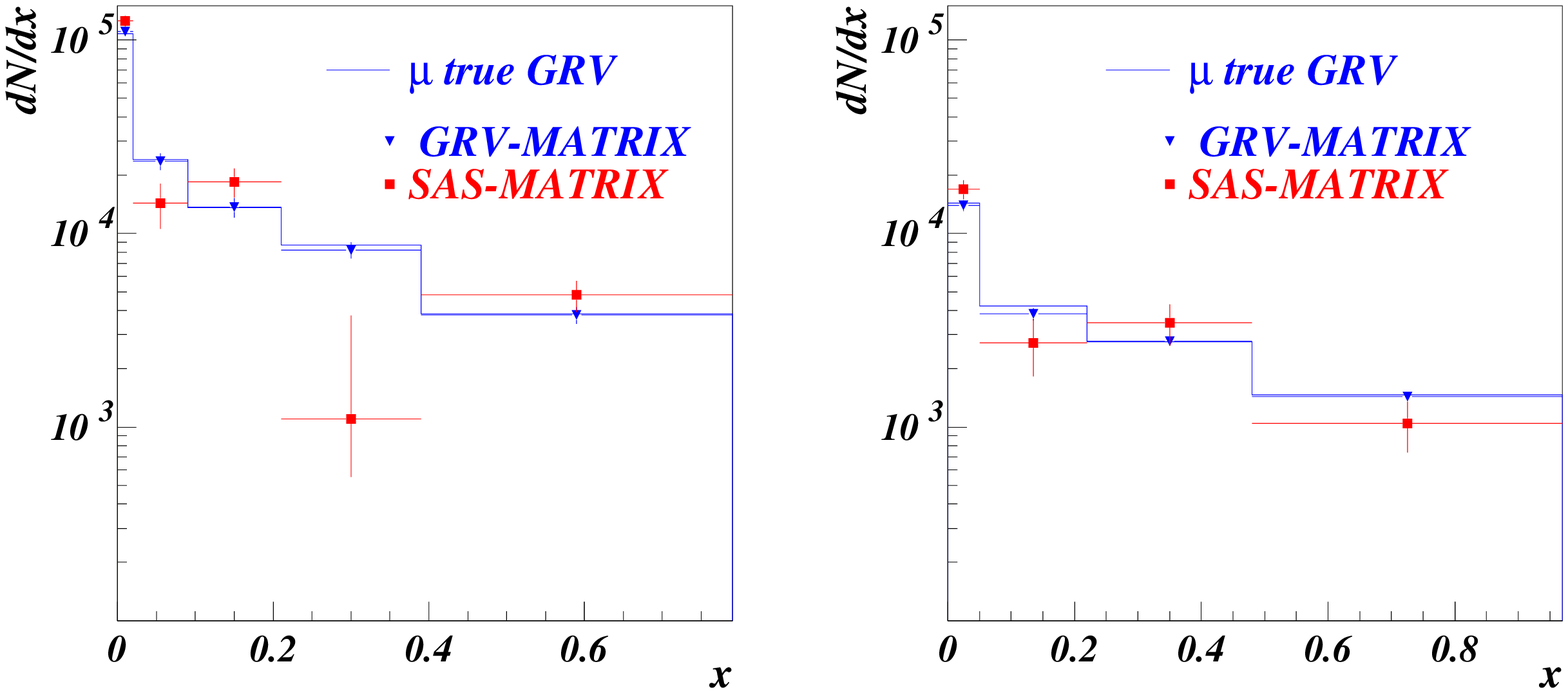,width=9.15cm}
\epsfig{file=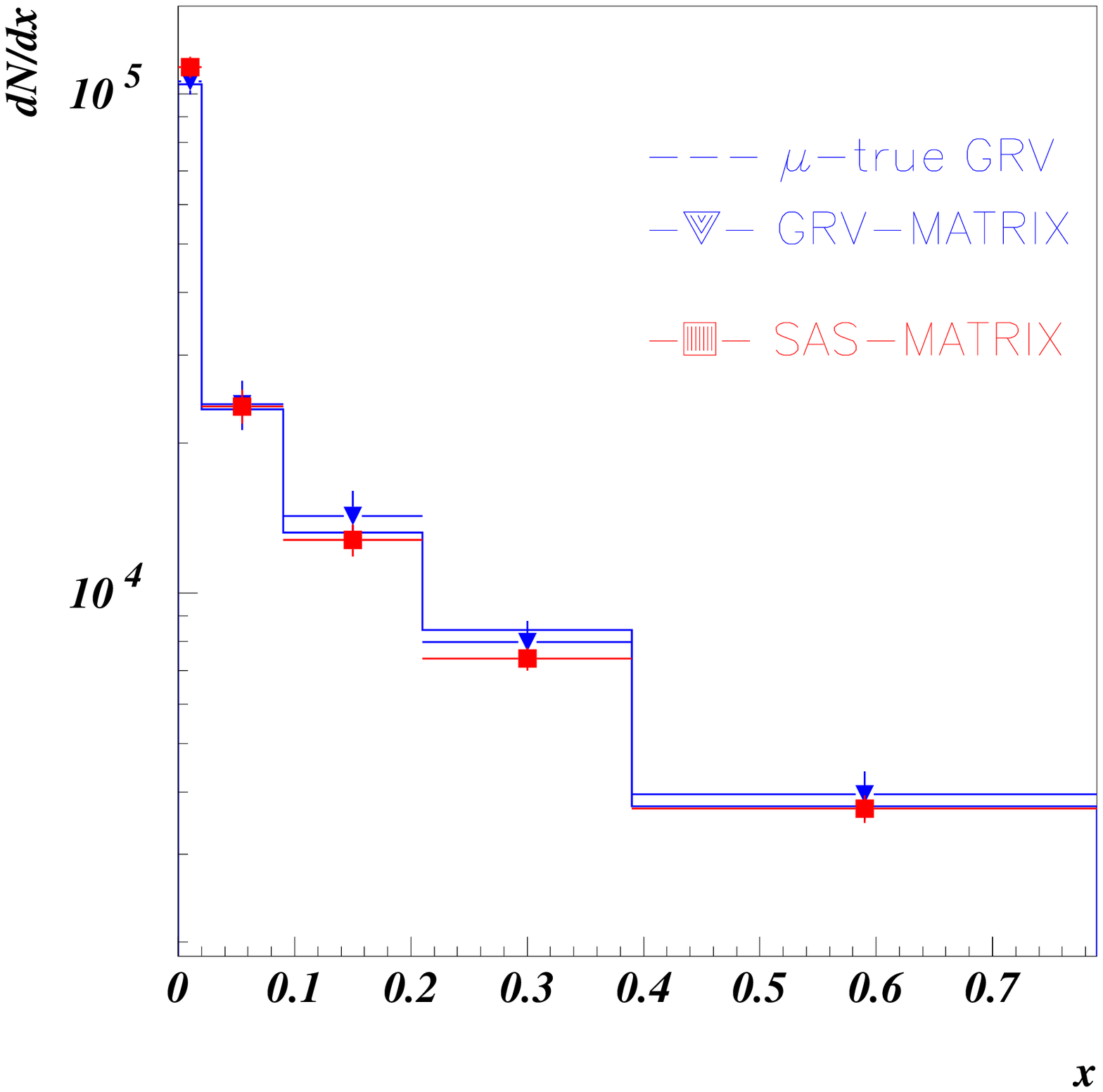,width=4.2cm}\nolinebreak\hspace*{-0.25cm}
\epsfig{file=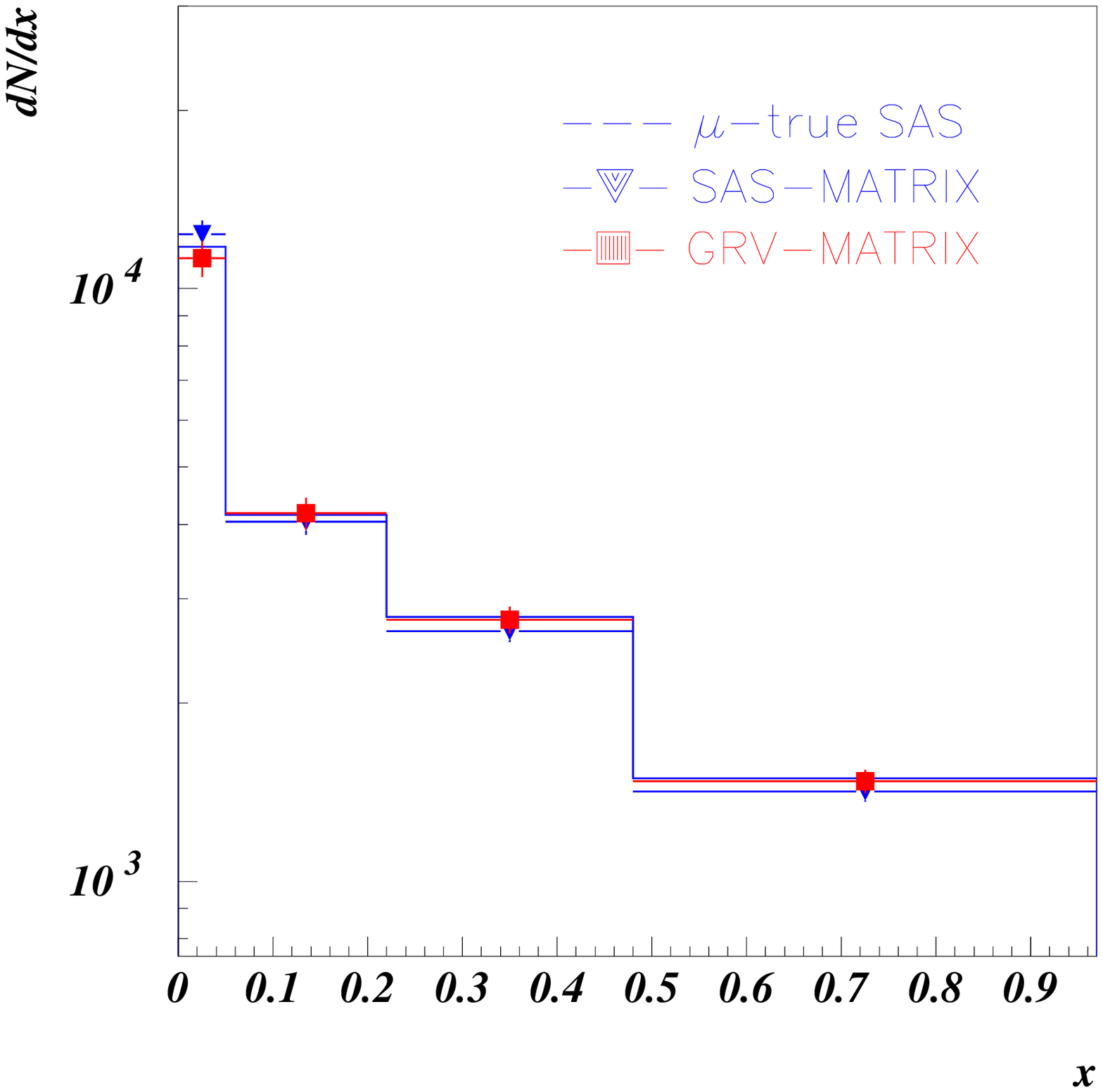,width=4.2cm}
\vspace*{-1.2cm}
\caption{One- (upper) and two-dimensional (lower plots) test 
unfolding in SiCAL- (left) and LCAL- 
(right) tagged events using a sample of toy-MC.}
\label{fig-dim12}
\end{figure}

The two-dimensional unfolding leads to smaller statistical errors. This 
is because in the one-dimensional case, the effective weight of each event 
is determined by the average $x$ resolution in the bin. With 
two-dimensional unfolding, those events with low $E_{17}$ are given 
a higher weight in the final result. The improvements achievable by 
two-dimensional unfolding where investigated quantitatively in 
Ref.~\cite{cartwright}. An example is shown in Fig.~\ref{fig-dim12}, 
where a toy-MC sample was used and two different response matrices 
to demonstrate the differences in one- 
and two-dimensional unfolding. Nevertheless the model dependence remains a 
significant source of systematic uncertainty, and several event generators 
are used in the presented analysis to account for this. After unfolding the 
structure function $F_2^{\gamma}$ is obtained from the $x$ distribution 
${\mathrm d}N/{\mathrm d}x$ using the same MC simulation that was used for 
the determination of the response matrix. The measurement in data is taken as 
the average of the distributions obtained with the response matrices of the 
models used.

\vspace*{-0.8cm}
\vspace*{-0.8cm}

\section{RESULTS}

\subsection{Measurements for $F_2^{\gamma}$}

\begin{figure}[htb]
\epsfig{file=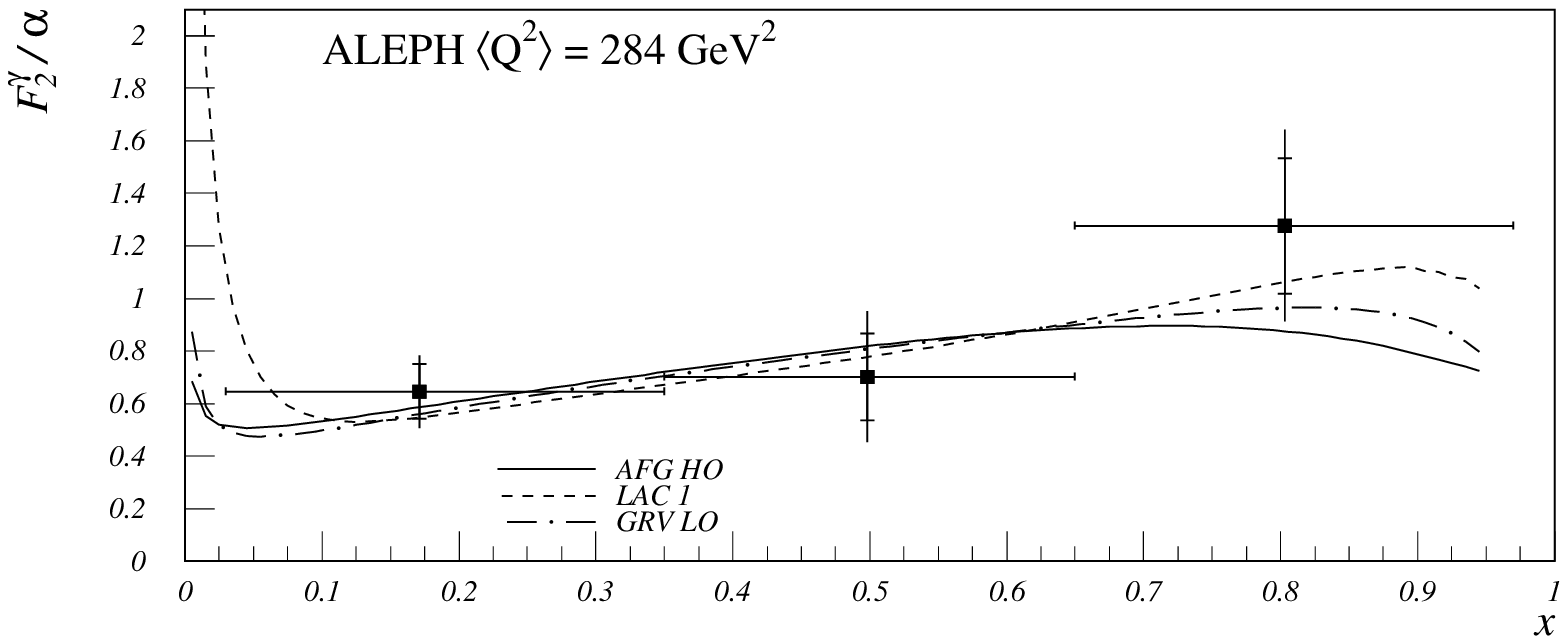,width=7.5cm}
\epsfig{file=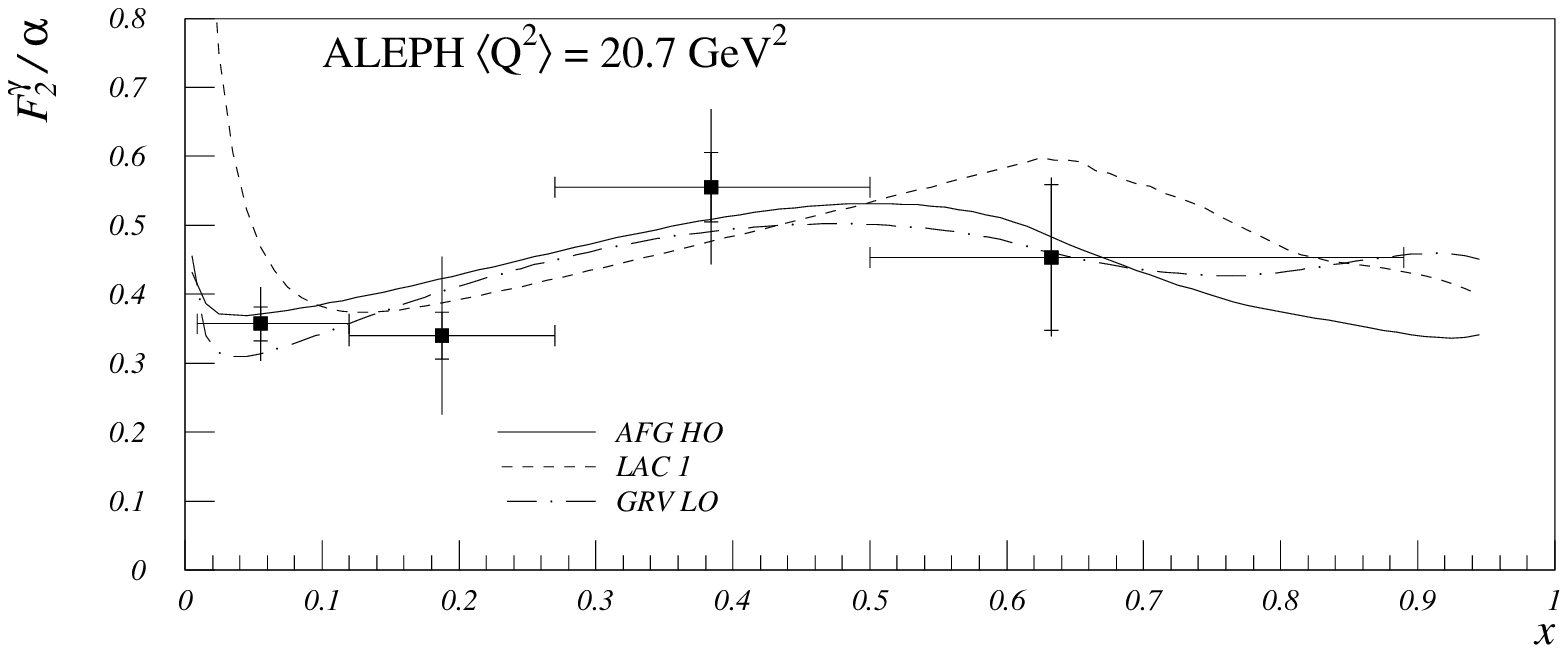,width=7.5cm}
\epsfig{file=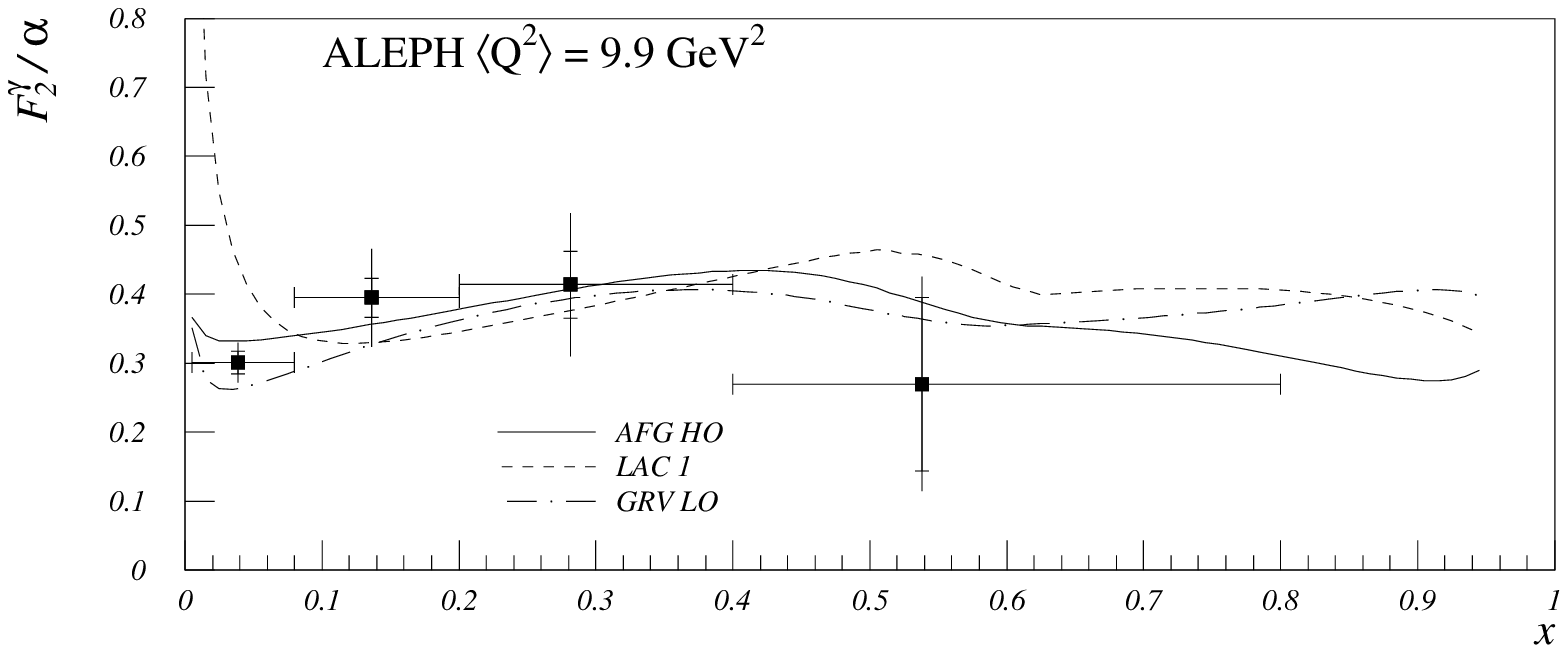,width=7.5cm}
\caption{The values for $F_2^{\gamma}/\alpha$ at 
$\sqrt{s} = 91\,{\mathrm{GeV}}$ 
compared to three different parameterizations.}
\label{fig-f2gamma91}
\end{figure}

\begin{figure}[htb]
\hspace*{-0.55cm}
\epsfig{file=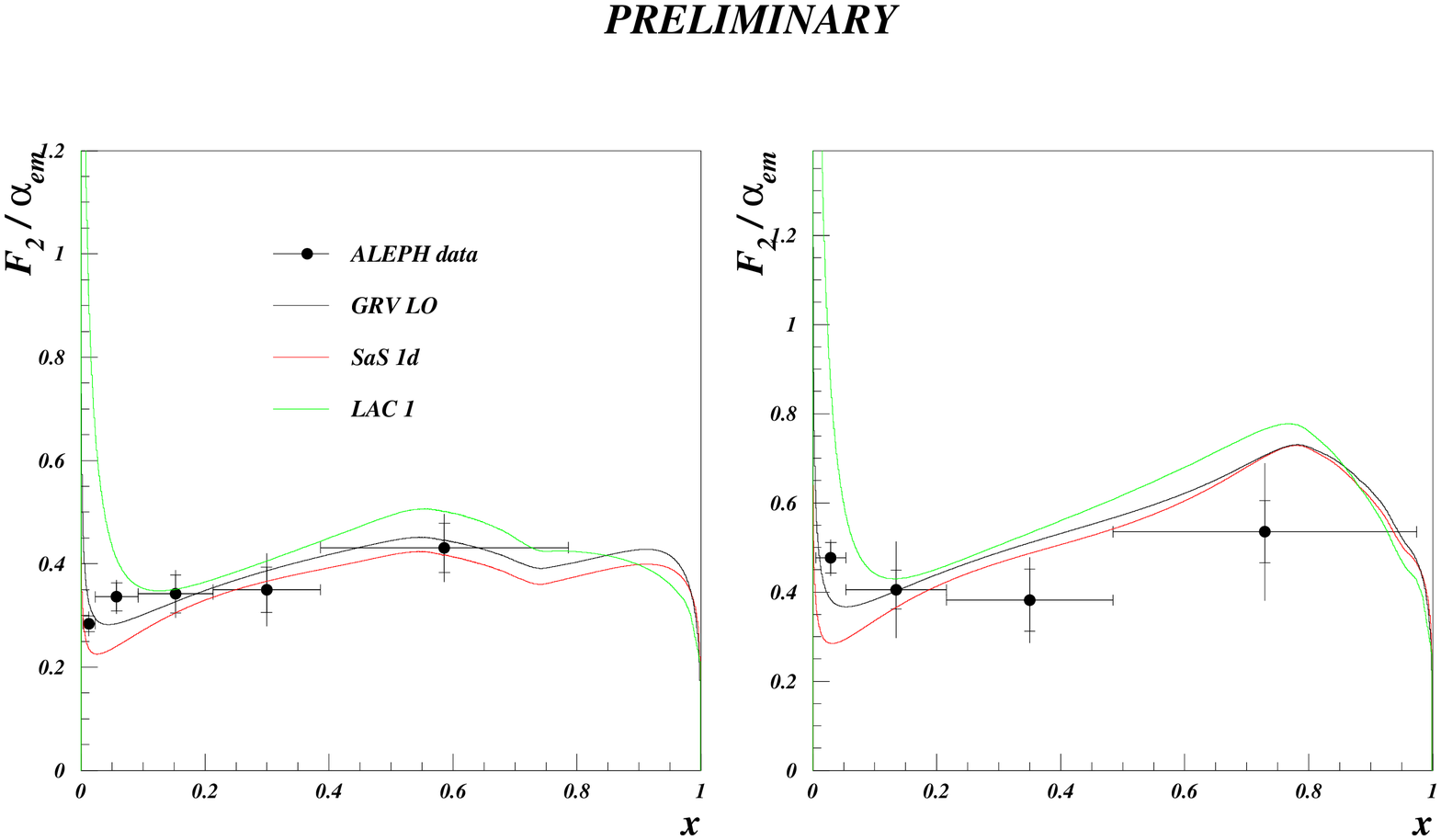,width=8.8cm}
\vspace*{-1.2cm}
\caption{The values for $F_2^{\gamma}/\alpha$ at 
$\sqrt{s} = 183\,{\mathrm{GeV}}$ 
($\langle Q^2 \rangle$ =13.7 and $56.5\,{\mathrm{GeV^2}}$) 
compared to three different parameterizations.}
\label{fig-f2gamma183}
\end{figure}

The final results for $F_2^{\gamma}$ are shown in Fig.~\ref{fig-f2gamma91} 
for $\sqrt{s} = 91\,{\mathrm{GeV}}$ and in Fig.~\ref{fig-f2gamma183} for 
$\sqrt{s} = 183\,{\mathrm{GeV}}$, resp. The inner error bars are statistical 
errors, the total error bars represent the quadratic sum of statistical 
and systematic errors. The systematic error comprise uncertainties from 
the unfolding (spread of the results based on the different models etc.), 
from detector efficiency and acceptance including trigger, 
selection criteria etc. The $91\,{\mathrm{GeV}}$ analysis includes a small 
theoretical error from the assumed form factor of the virtual photon as well 
as the dependence on the assumption of $F_L^{\gamma}$.

\vspace*{-0.8cm}

\subsection{Comparison of $F_2^{\gamma}$ to parameterizations}

For both centre-of-mass energies and the different $Q^2$ ranges all of the 
parameterizations provide a good description of the data for $x > 0.1$. 
For lower $x$ the LAC values are significantly too high. At 
$\sqrt{s} = 91\,{\mathrm{GeV}}$ a test has been made using 
event reweighting to ensure that the difference is not an 
artifact of the unfolding procedure; for more details see 
Ref.~\cite{f2gamma91}. The low $x$ region is sensitive to the 
gluon content of the photon. A $\chi^2$ was calculated in the eleven bins in 
$x$ to quantify the comparison to various sets of the parton density 
function of the photon, see Table \ref{tab-chi2}. Those that show 
significantly large values of $\chi^2$ such as LAC 1 and 2 and Whit 4,5, and 
6, contain a large gluon content, resulting in a rapid rise in the 
structure function at low $x$.

\begin{table}[hbt]
\caption{The values of $\chi^2$ obtained data at to the more 
recently calculated photon parton density function, see 
\protect\cite{f2gamma91} and references therein.}
\label{tab-chi2}
\begin{tabular}{|l|r|l|r|}
\hline
\multicolumn{1}{|c}{PDF} & \multicolumn{1}{|c|}{$\chi^2$} & 
\multicolumn{1}{|c}{PDF} & \multicolumn{1}{|c|}{$\chi^2$} \\
\hline
DG Set 1 &   4.3  & DG Set 2 &   5.0 \\
LAC 1    & 107.2  & LAC 2    &  75.9 \\
LAC 3    &   3.9  & GS-96 HO &   7.6 \\
GS-96 LO &   8.9  & GRV HO   &   4.9 \\
GRV LO   &   3.8  & AFG HO   &   4.6 \\
Whit 1   &   5.2  & Whit 2   &  13.9 \\
Whit 3   &  18.3  & Whit 4   &  40.0 \\
Whit 5   & 105.5  & Whit 6   & 130.8 \\
SaS Set 1D & 10.0 & SaS Set 1M & 4.3 \\
SaS Set 2D & 3.6  & SaS Set 2M & 3.7 \\
\hline
\end{tabular}
\end{table}

\subsection{Dependence of $F_2^{\gamma}$ from $\langle Q^2 \rangle$}

The values of $F_2^{\gamma}$ for the three $Q^2$ ranges at 
$\sqrt{s} = 91\,{\mathrm{GeV}}$ have been averaged over the region 
$0.1 < x < 0.6$. A logarithmic rise with $\langle Q^2 \rangle$ is 
seen as expected from theoretical predictions using the 
parameterizations listed in Table \ref{tab-chi2}.

\section{CONCLUSIONS}

Single-tagged two-photon events recorded by the ALEPH detector at LEP I and 
LEP II have been studied in three and two bins of $Q^2$, which have a mean of 
9.9, 20.7 and $284\,{\mathrm{GeV^2}}$ and 13.7 and $56.5\,{\mathrm{GeV^2}}$. 
The data have been used to measure the hadronic structure function 
$F_2^{\gamma}$ as a function of $x$. The comparison with parameterizations 
show that those with parton density functions that contain a large gluon 
content are inconsistent with data. The rise of $F_2^{\gamma}$ with 
$\langle Q^2 \rangle$ is found compatible with the available parameterizations. 

A two-dimensional unfolding technique using the principle of maximum 
entropy has been successfully applied; with the second variable chosen as 
$E_{17}$, defined as the total energy of the particles having 
angles with respect to the beam line less than $17^{\circ}$. This 
unfolding method reduces the 
statistical errors and the model dependence of the extracted $F_2^{\gamma}$ 
as compared to one-dimensional procedures.

\end{document}